\documentclass[%
 reprint,
 amsmath,amssymb,
 aps,
prc,
]{revtex4-1}

\usepackage{graphicx}% Include figure files
\usepackage{dcolumn}% Align table columns on decimal point
\usepackage{bm}% bold math
\usepackage{hyperref}% add hypertext capabilities
\usepackage{xcolor}

\begin{document}

%\preprint{APS/123-QED}

\title{Experimental study of the $^{17}$F+$^{12}$C fusion reaction and its implications for fusion of proton-halo systems }

\author{B. W. Asher}
\email[Email address: ]{bwa15@my.fsu.edu}
\author{S. Almaraz-Calderon}
\email[Email address: ]{salmarazcalderon@fsu.edu}
\author{Vandana Tripathi}
\author{K. W. Kemper}
\author{L. T. Baby}
\author{N. Gerken}
\author{E. Lopez-Saavedra}
\author{A. B. Morelock}
\author{J. F. Perello}
\author{I. Wiedenh{\"o}ver}
\affiliation{Department of Physics, Florida State University, Tallahassee, Florida 32306, USA}
\author{N. Keeley}
\affiliation{National Centre for Nuclear Research, ul.\ Andrzeja So\l tana 7, 05-400
Otwock, Poland}

% It is always \today, today,
             %  but any date may be explicitly specified

\begin{abstract}
The total fusion cross section for the $^{17}$F + $^{12}$C system at incident energies near the top of the Coulomb barrier was studied using the newly developed \emph{Encore} active-target detector at Florida State University. The $^{17}$F nucleus exhibits interesting nuclear structure properties in that while it has a low threshold against $^{17}\mathrm{F} \rightarrow \protect{^{16}\mathrm{O}} + p$ breakup ($S_p = 600$ keV), the valence proton is in the 1d$_{5/2}$ shell in the ground state, so that the nuclear matter radius is predicted to be similar to that of the $^{16}$O core. By contrast, the low-lying $1/2^+$ first excited state ($E_\mathrm{ex}$ = 105 keV), with the valence proton in the 2s$_{1/2}$ shell, is considered to be a proton halo. In this work possible influences of both the weak binding and the halo nature of the excited state on the total fusion cross section were investigated. The new data reported here complement existing measurements for the total fusion of $^{17}$F with heavy ($^{208}$Pb) and medium mass ($^{58}$Ni) targets by extending the range of systems studied to one where Coulomb effects should be minimal. Total fusion cross sections for the stable counterpart systems $^{16}$O + $^{12}$C and $^{19}$F + $^{12}$C were also measured to enable a systematic comparison. No significant influence of either the weak binding or the halo nature of the $^{17}$F $1/2^+$ first excited state on the above barrier total fusion excitation function was observed when compared with the stable counterpart systems.  

\end{abstract}

\pacs{Valid PACS appear here}% PACS, the Physics and Astronomy
                             % Classification Scheme.
%Use showkeys class option if keyword
                              %display desired
\maketitle

%\tableofcontents

\section{Introduction}

Fusion measurements are a key component of research in nuclear structure, nuclear reactions and nuclear astrophysics \cite{BackReview}. For example, the $^{12}$C + $^{12}$C fusion reaction determines the burning conditions and subsequent isotopic composition of the resulting ashes in massive stars \cite{jiang12C}. Furthermore, in neutron-rich stars fusion of exotic carbon and oxygen isotopes may act as catalyzers for the so called X-ray superbursts \cite{CarnelliPrl}. 
Recently, fusion reaction experiments involving light exotic beams have become the focus of several studies since such nuclei have become accessible at existing facilities. This area of research will only grow with the forthcoming exotic beam facilities around the world. Beams of short-lived radioactive nuclei present unique opportunities to probe the dynamics of fusion reactions around the Coulomb barrier. Weakly-bound light exotic nuclei in particular provide the possibility to explore the interplay between fusion, breakup and transfer reactions over a much wider range of binding energies and structural properties than those available with stable beams \cite{keeley07}. A specific sub-class of this type of nucleus are the so-called halo nuclei which have extended matter distributions \cite{tanihata,tanihatahe} such that large breakup and/or transfer cross sections are observed at incident energies close to the Coulomb barrier \cite{canto15}. 

It has long been suggested that the fusion cross section should be significantly enhanced in systems involving halo nuclei \cite{balantekin98, ErnstFlourine} due to their extended size, since the fusion probability is highly dependent on the size and the shape of the interacting nuclei. On the other hand, due to their low threshold against breakup it has also been suggested that there could be significant suppression of fusion in such systems. However, the question of whether or not fusion in systems involving halo nuclei is enhanced has not yet been satisfactorily answered experimentally. In fact, fusion reactions involving halo nuclei have led to contradictory conclusions \cite{ErnstFlourine, balantekin98, keeley07, kolatareview} which could be the result of experimental uncertainties but also the lack of sufficient data for a systematic comparison between systems. A more fundamental problem is the lack of general agreement as to the benchmark used to infer enhancement or suppression and whether complete fusion or total fusion should be considered (see, e.g., Ref.\ \cite{keeley07} for a discussion of these questions).

Most experiments with halo nuclei have been carried out on the neutron-rich side of the chart of the nuclides \cite{keeley07}. Results of fusion reactions with the neutron-rich halo nuclei  $^{6,8}$He \cite{6He,8He}, $^{11}$Li \cite{11Li} and $^{11}$Be \cite{11Be,11Beanalysis} show an effect on the fusion excitation function which is mainly manifested as a reduction in the cross section above the barrier. This effect has been explained by the low neutron removal thresholds of these systems \cite{kolatareview}, resulting in the loss of beam flux at relatively large distances between the colliding nuclei due to breakup itself and/or neutron transfer reactions.

For proton-halo nuclei, despite the expanded size of the halo their weakly-bound nature might also be expected to manifest itself as a reduction of the fusion cross section above the Coulomb barrier \cite{kolatareview,halostuff}. However, the few available experimental results appear inconsistent. For example, the proton-halo nucleus $^{8}$B has been the object of various studies with $^{58}$Ni and $^{28}$Si \cite{BorNi,BorSi} targets. The $^{8}$B + $^{58}$Ni system showed an enhancement in the fusion cross section in all regions including well above the Coulomb barrier while the $^{8}$B + $^{28}$Si system shows a slight suppression above the barrier \cite{BorNi,BorSi}.

The conclusions concerning the fusion of $^{17}$F, the focus of this work, are also not definitive. The $^{17}$F nucleus has a low breakup threshold ($S_p$ = 600 keV) but since its ground state is usually deemed to consist of a proton in the 1d$_{5/2}$ shell outside the doubly magic $^{16}$O core it is not considered to constitute a halo due to the large centrifugal barrier. Rather, it is the low-lying $1/2^+$ first excited state ($E_\mathrm{ex}$ = 105 keV) with the ``valence'' proton in the 2s$_{1/2}$ shell which is thought to be a proton halo \cite{17FHalo}. The inherent nuclear structure properties of $^{17}$F therefore make it a prime candidate for reaction studies and for investigating the effect on the fusion cross section of a possible proton halo in a low-lying bound excited state rather than the ground state. 

There have been several experimental studies using $^{17}$F beams \cite{ErnstFlourine,FlNi,FlPb,flourinereaction,Mazz2010,Liang2000,Lian2003,Sig2010,Romoli2004,Yang2021}. Among these one study measured the fusion-fission cross section for a $^{208}$Pb target where it was concluded that at energies around the Coulomb barrier no enhancement of the fusion cross section is observed compared to those for the stable $^{19}$F and the $^{16}$O core with the same target \cite{ErnstFlourine}, in contrast with the results for the $^{8}$B + $^{58}$Ni system \cite{BorNi} but consistent with those for $^8$B + $^{28}$Si \cite{BorSi}. It has been suggested that this lack of an enhancement in the fusion cross section could be due to an effective polarization of the $^{17}$F in the strong Coulomb field of the $^{208}$Pb target, leading to a shielding effect on the halo proton \cite{shielding}. In a recent experiment where the reaction dynamics of the $^{17}$F + $^{58}$Ni system was studied, it was found that the behavior of the total fusion cross section was identical with that of the $^{16}$O + $^{58}$Ni system at above-barrier energies, but enhancement was observed below the Coulomb barrier which coupled discretized continuum channels (CDCC) calculations demonstrated was due to the effect of couplings to the $^{17}\mathrm{F} \rightarrow \protect{^{16}\mathrm{O}} + p$ breakup process \cite{Yang2021}.

The present work reports a measurement of the total fusion cross section excitation function for the $^{17}$F + $^{12}$C system at energies around the Coulomb barrier to search for the effects of its weak binding in a light mass system, thus obviating any possible shielding effects and complementing the existing data sets for medium and heavy mass targets by extending the range of studies to a system where Coulomb effects should be minimal. A novel detector system developed at Florida State University allows for simultaneous detection of the incoming beam and the fusion products. The fusion products are measured simultaneously over an extended energy range without changing the energy of the incoming beam. The same experimental conditions were used to measure the fusion cross sections for the more tightly-bound stable systems $^{16}$O + $^{12}$C and $^{19}$F + $^{12}$C at energies near the Coulomb barrier, thus enabling a direct comparison of the fusion cross sections for all three systems.

\section{Experimental Details}

The experiment was performed at the John D. Fox accelerator laboratory at Florida State University (FSU). A $^{17}$F radioactive beam was produced by the RESOLUT radioactive beam facility \cite{resolut}. A stable $^{16}$O beam from the SNICS ion source was accelerated to 64.5 MeV by the tandem Van de Graaff accelerator and boosted to 91.5 MeV by the coupled LINAC accelerator. A liquid nitrogen cooled deuterium gas production target kept at a pressure of 350 torr was bombarded with the $^{16}$O beam. The radioactive $^{17}$F beam (t$_{1/2}$ = 64.5 s) was produced at a rate of $\sim$ 600 particles per second (pps) in-flight via the $^{16}$O(d,n)$^{17}$F reaction and focused onto the detector system by the super conducting solenoid of RESOLUT. The main contaminant was the primary $^{16}$O beam at a rate of $\sim$ 1100 pps, which was used simultaneously in our experiment. A measurement with a stable $^{19}$F beam from the Tandem accelerator was also performed. 

The 69.1 MeV $^{17}$F beam and its main contaminant, the 58.1 MeV $^{16}$O beam, were delivered to the \emph{Encore} active-target detector. \emph{Encore} is a multi-sampling ionization chamber recently developed at FSU, optimized to measure fusion cross sections with low-intensity exotic beams ($\leq 10$ kHz). \emph{Encore} is based on the MUSIC detector at Argonne National Laboratory (ANL) \cite{CarnelliNim}. Details of the \emph{Encore} detector will be published in a separate paper \cite{EncoreNim}. This detector system and analysis procedure has been successfully used at ANL for measurements of fusion reactions with carbon isotopes \cite{CarnelliPrl} as well as for measurements of ($\alpha$,p) and ($\alpha$,n) reactions \cite{AvilaPrc, rashi}.

\begin{figure}
\centering
\includegraphics[width=\linewidth]{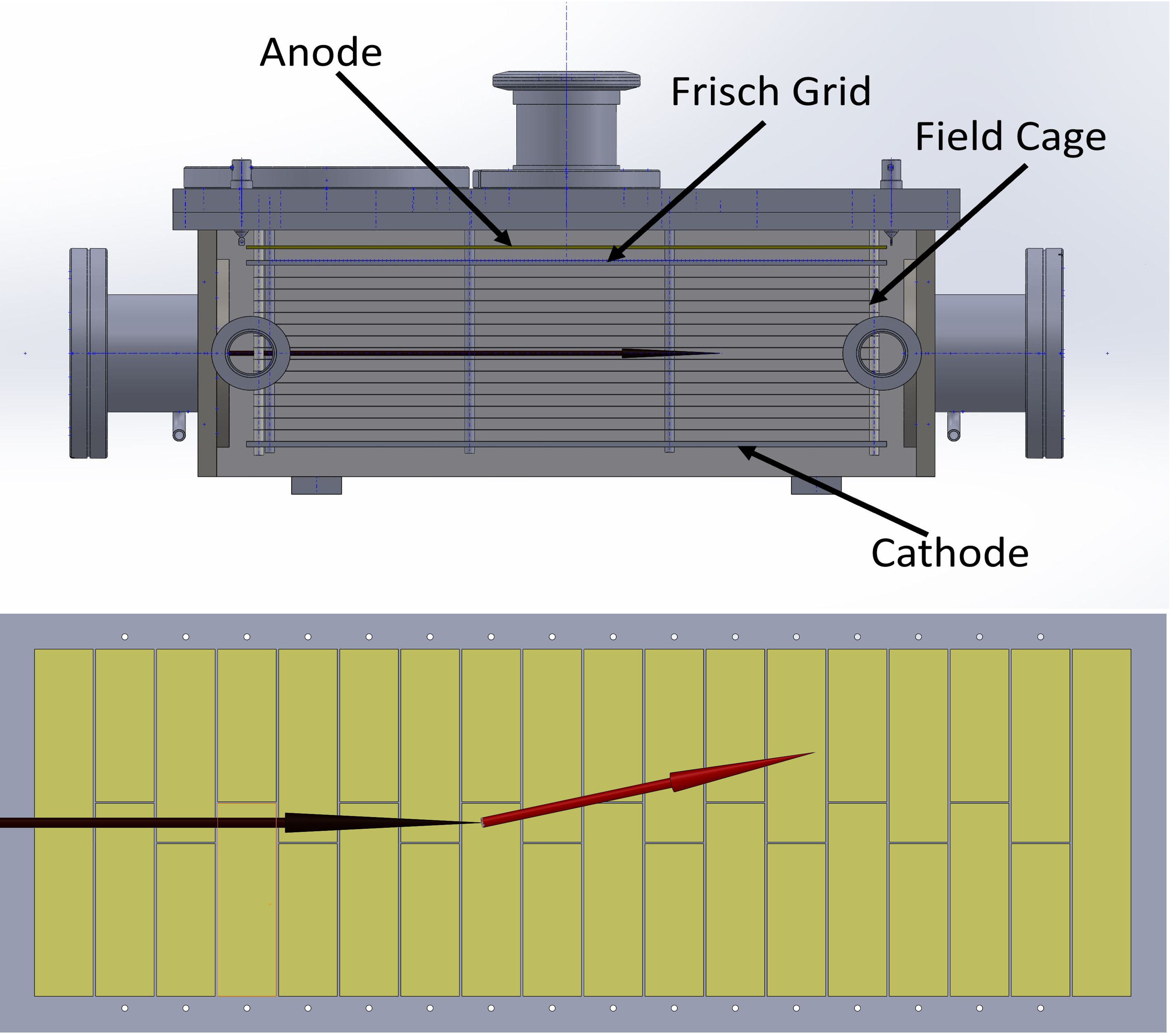}
\caption{Upper panel: 3D schematic of the \textit{Encore} detector, a multi sampling ionization chamber where the field cage produces a perpendicular electric field. The field cage consists of a negatively biased cathode, voltage divider wired planes, a Frisch grid and a segmented anode. The beam passes through the center of the active region. Lower panel: Schematic view of the segmented anode. The first and last strips are used as veto and control, respectively. The 16 strips subdivided into left and right halves are also shown. The black arrow indicates a beam particle entering the active region of the detector. A fusion reaction occurs in strip 7 creating an evaporation residue (red arrow) which is identified by its larger energy loss signal.}
\label{fig:encore1}
\end{figure}

A schematic view of the \emph{Encore} detector is shown in the upper panel of Fig.\ \ref{fig:encore1}. The beam enters through a 2.11 mg/cm$^{2}$ HAVAR window. \emph{Encore} works as an ionization chamber with an electric field perpendicular to the beam axis. The detector is filled with gas which serves as both target and counting material. \emph{Encore} measures energy losses as the beam passes through the detector. Ionization electrons produced by the interactions of the beam with the gas drift towards the segmented anode where the charge is collected, providing a signal proportional to the energy deposited by the ionizing particle.

After the beam enters the detector it travels 3 cm in a dead region before entering the segmented anode region. Energy losses of the beam are measured as it passes through the detector via 16 anode signals (strips 1, 2, \dots, 16) subdivided into left and right as shown in the lower panel of Fig.\ \ref{fig:encore1}. The left and right halves of the 16 strips are independently read using two 16-channel MPR-16 pre-amplifiers connected with high density FGG lemo cables and added together in the analysis. Two extra anode signals at the beginning and end of the detector (strips 0 and 17) are read individually and are used for vetoing and control. Signals from the cathode and the Frisch grid are also read out. For this experiment \emph{Encore} was filled with CH$_4$ gas at 168 torr. A gas handling system was used to re-circulate the gas inside the detector. The pressure of the gas was constantly monitored by a pressure gauge. The value of the pressure of the gas in the detector remained constant within the precision of the gauge meter (0.5\%) during the full measurement. The CH$_4$ gas used contains less than 1\% of $^{13}$C, therefore any contribution from reactions with the $^{13}$C in the gas is negligible. The gain of the anode segments was optimized to be more sensitive to signals corresponding to the interaction of the beam with the carbon in the CH$_4$ gas. Given the large difference in energy signals, the detector was not sensitive to the light particles or to interactions between the beam and the hydrogen in the gas.

\begin{figure}
\centering
\includegraphics[width=\linewidth]{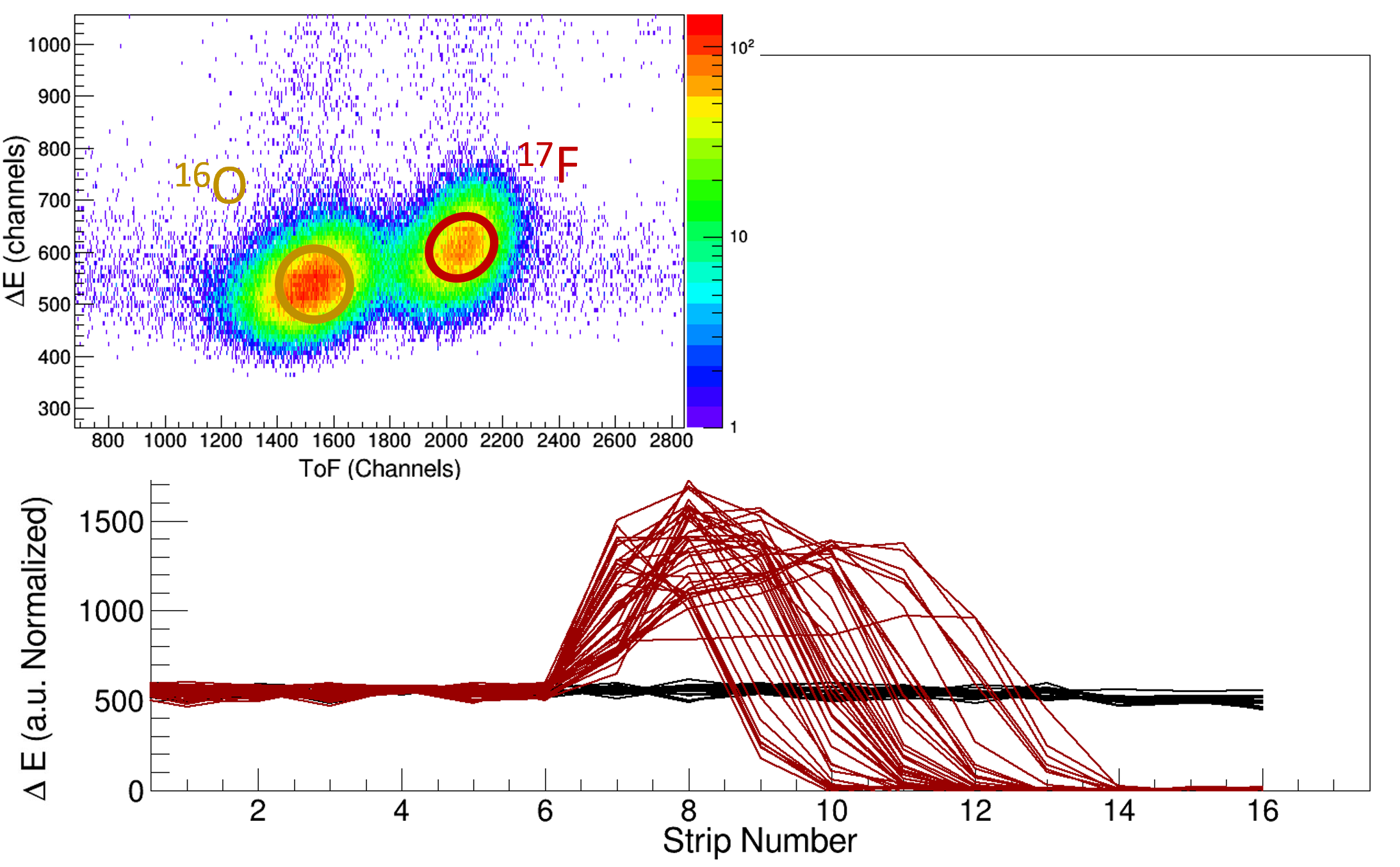}
\caption{Experimental traces measured in \textit{Encore}. The beam traces inside the detector (black) have been normalized to a fixed value for the analysis. The normalization allows a consistent threshold to be set when searching for energy jumps within the segmented anode strips of the detector. Experimental $^{17}$F + $^{12}$C fusion events occurring in strip 7 are shown by the red traces. A jump in energy loss is seen as a result of the creation of an evaporation residue which will stop in the detector prior to the beam. The inset in the top-left corner shows the separation of the $^{17}$F and $^{16}$O beams in \emph{Encore} due to their different time-of-flight.}
\label{fig:encore2}
\end{figure}

The energy losses measured in each strip are analyzed on an event-by-event basis. One event through the detector, composed of 16 right side anode, 16 left side anode, 1 strip 0, and 1 strip 17 signals, is called a trace. Most of the time \emph{Encore} measures beam-like events. A sample of the experimental $^{17}$F beam traces is shown by the black lines in Fig. \ref{fig:encore2} where they are normalized to channel 500 for the analysis.

Guided by energy loss simulations, an algorithm was developed to search for fusion reactions in the detector on an event-by-event basis. Fusion-like events are characterized by a beam-like trace followed by a sudden jump in the energy loss in the specific strip where the fusion reaction occurs due to the larger charge of the evaporation residue. The evaporation residue loses much more energy than the beam, therefore the fusion trace stays high for a few strips before going to zero. Experimental fusion traces for the $^{17}$F + $^{12}$C system occurring in strip 7 are shown by the red lines in Fig.\ \ref{fig:encore2}.  
The left and right segmentation of the anode provides multiplicity information which allows events like elastic and inelastic scattering to be identified. Fusion events and beam events happen in either the right or the left side of the anode strips (multiplicity one), while scattering events have signals in both sides of the anode strips (multiplicity two) and can be easily rejected.

\emph{Encore} provides full angular coverage of the evaporation residues allowing for a measurement of the total fusion cross section per strip. This translates into a measurement of the fusion excitation function of the system over a wide energy range using a single beam energy. The range of the excitation function is determined solely by the energy of the beam and the gas pressure in the detector. Since \emph{Encore} measures all the beam all the time, it provides an absolute beam normalization of the measured total fusion cross sections. \emph{Encore} is particularly efficient for fusion measurements with low-intensity exotic beams ($\leq 10$ kHz) since there is no need to re-tune the energy of the beam to measure several points in an excitation function. 

\section{Experimental Results}

A silicon detector was mounted inside, at the end of the detector, for beam tuning purposes. Without gas in the detector and after the entrance HAVAR window, the energy of the $^{17}$F beam was measured to be 61.2 MeV (FWHM = 2 MeV) in the laboratory frame while that of the primary $^{16}$O$^{8+}$ beam was measured to be 51.2 MeV (FWHM = 1.7 MeV). Once the detector was filled with gas, the $^{17}$F and $^{16}$O beams were separated in \emph{Encore} by their different time-of-flight and their different $\Delta$E signals as shown in the inset to Fig. \ref{fig:encore2}.

In the present experiment the fusion excitation function of the $^{17}$F + $^{12}$C system was measured inside the active region of the detector over the range in center of mass energy corresponding to $E_\mathrm{c.m.}$ = 19.4 - 9.0 MeV, with an average energy of 1.2 MeV deposited in each strip. In order to extract the total fusion cross section per strip, the identified fusion events in a given strip are counted and normalized by the number of beam events in the detector. The corresponding energy and target thickness per strip, determined by the beam energy, gas pressure inside the detector and size of the strip, were calculated using LISE++ \cite{lise}. The study by Carnelli {\it et al.} \cite{CarnelliNim} showed the validity of this approach. The error bars on the cross section measurements are dominated by statistics. Systematic uncertainties are due to target thickness (i.e., the size of the anode strips, the pressure of the gas, variations in temperature). A conservative minimum of 10\% error bars on the cross sections has been adopted to account for systematic uncertainties.
The error bars in energy arise from the 1.5 cm thickness of the strips and the possibility of the reaction occurring anywhere within the width of a particular strip. The error bars in the energy consider that the reaction occurred in the middle of the strip. The measured total fusion cross section values are reported in Table \ref{table:table1}.

\begin{table}
\begin{center}
\begin{tabular}{ |c|c| } 
 \hline
 $E_\mathrm{c.m.}$ (MeV) & $\sigma$ (mb)   \\ 
 \hline
19.4 $\pm$ 0.5 & 914 $\pm$ 88	\\
18.4 $\pm$ 0.5 & 797 $\pm$ 89	\\
17.4 $\pm$ 0.5 & 728 $\pm$ 85	\\
16.3 $\pm$ 0.5 & 789 $\pm$ 91 \\
15.2 $\pm$ 0.5 & 831 $\pm$ 94	\\
14.0 $\pm$ 0.6 & 704 $\pm$ 89	\\
12.8 $\pm$ 0.6 & 660 $\pm$ 86	\\
11.6 $\pm$ 0.6 & 516 $\pm$ 81  \\
10.3 $\pm$ 0.6 & 240 $\pm$ 55	\\
8.9 $\pm$ 0.7 & 150 $\pm$ 47	\\
 \hline
\end{tabular}
\caption{Total fusion cross sections for the $^{17}$F + $^{12}$C system measured in the present experiment as a function of center of mass energy.}
\label{table:table1}
\end{center}
\end{table}

\begin{figure}
\centering
\includegraphics[width=\linewidth]{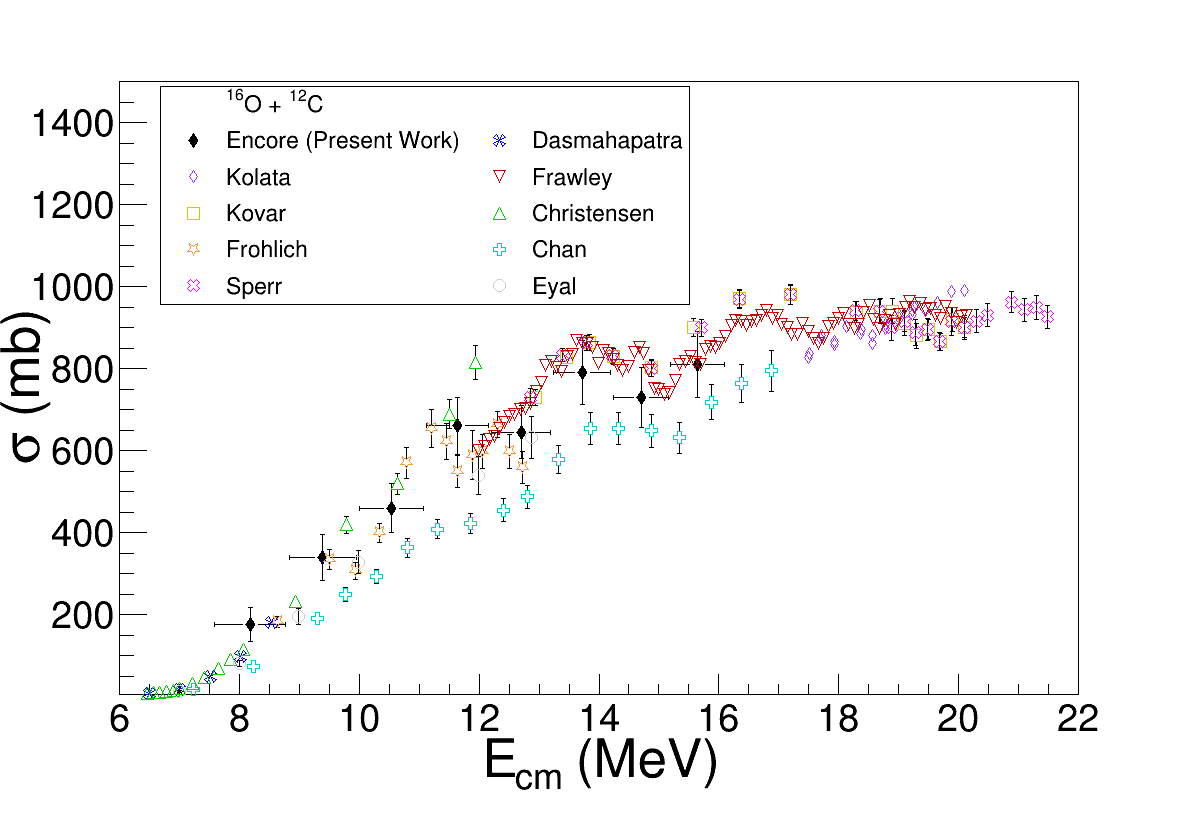}
\caption{Total fusion cross sections for $^{16}$O + $^{12}$C (black diamonds) measured in the present experiment compared with previously published data \cite{kovar16o,kolata16o,frohlich16o,chan16o,Sperr16o,Das16o,Fraw16o,Chri16o,Eyal16o}.}
%, in agreement with \cite{ErnstFlourine,flourinereaction}.}
\label{fig:16O16O}
\end{figure}

Fusion events from the primary $^{16}$O beam were measured simultaneously in \emph{Encore} with those for $^{17}$F. 
The fusion cross sections for the $^{16}$O + $^{12}$C system thus obtained are plotted on Fig. \ref{fig:16O16O} together with previous measurements from the literature \cite{kovar16o,kolata16o,frohlich16o,chan16o,Sperr16o,Das16o,Fraw16o,Chri16o,Eyal16o}. The good agreement between the present $^{16}$O + $^{12}$C fusion data and the previous measurements gives confidence in the $^{17}$F + $^{12}$C fusion measurements.

\begin{figure}
\centering
\includegraphics[width=\linewidth]{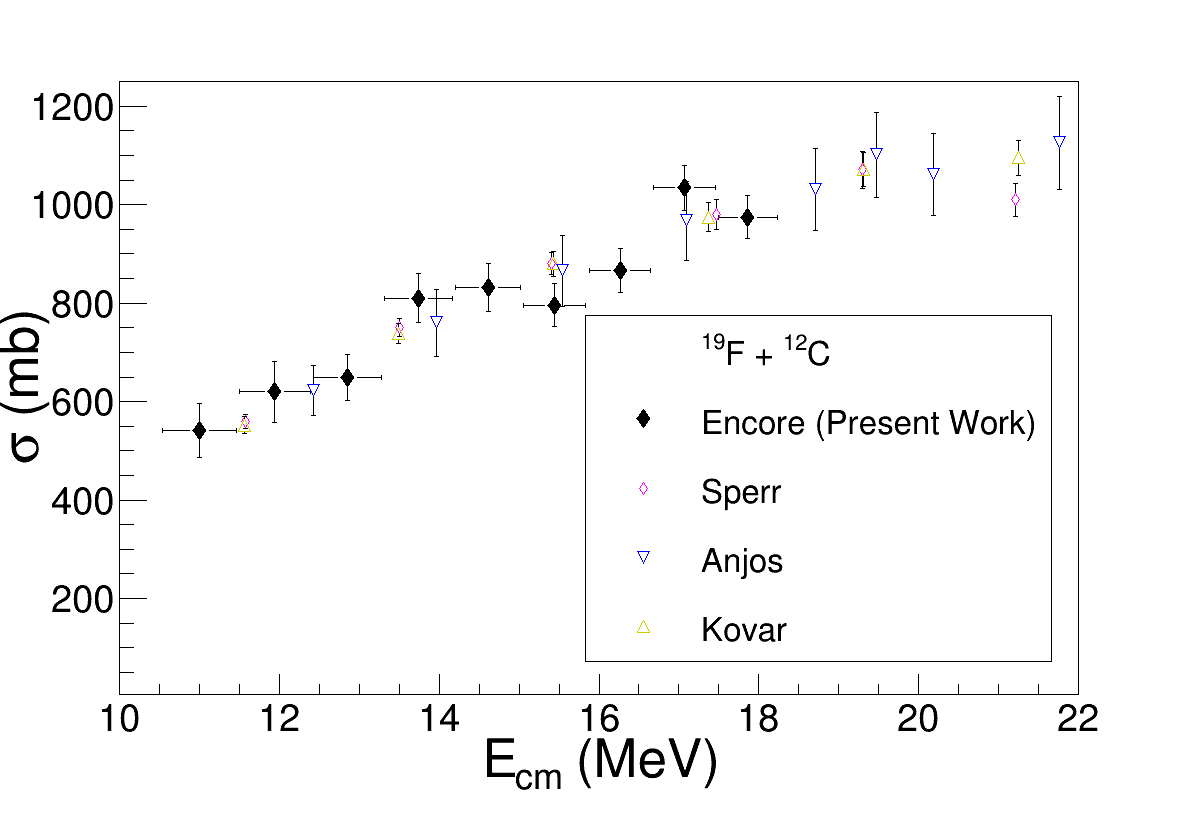}
\caption{Total fusion cross sections for $^{19}$F + $^{12}$C (black diamonds) measured in the present experiment compared with previously published data \cite{Kovar,Anjos,Sperr19F}.}
%, in agreement with \cite{ErnstFlourine,flourinereaction}.}
\label{fig:19F16O}
\end{figure}

In order to make a systematic comparison of any effects on the fusion cross section due to the exotic nature of $^{17}$F, we also performed a measurement with its stable counterpart $^{19}$F on a $^{12}$C target. A 65 MeV $^{19}$F beam at a rate of $\sim 1 \times 10^4$ pps was delivered to \textit{Encore} which was filled with  CH$_4$ gas at a pressure of 131 Torr. This pressure was chosen to scan a similar range in center of mass energy to the $^{17}$F + $^{12}$C measurement. The $^{19}$F arrived in the first control strip at 51 MeV, depositing between 0.75 MeV and 1.2 MeV in each strip, with an average of 0.9 MeV. The absolute cross sections for the $^{19}$F + $^{12}$C system in the energy range $E_\mathrm{c.m.}$ = 17.8 - 11 MeV measured in this experiment are plotted on Fig.\ \ref{fig:19F16O} together with previously published data \cite{Kovar,Anjos,Sperr19F}. The good agreement between the \emph{Encore} measurements and the previous data confirms the consistency of our analysis procedure.

%\begin{figure}
%\centering
%\includegraphics[width=\linewidth]{19F16OStandard-Oct.png}
%\caption{Total fusion cross sections for $^{16}$O + $^{12}$C (blue diamonds) and $^{19}$F + %$^{12}$C (red triangles) systems measured in the present experiment compared with %previously published data for both systems showing the good agreement obtained.}
%, in agreement with \cite{ErnstFlourine,flourinereaction}.}
%\label{fig:19F16O}
%\end{figure}

\section{Analysis}

One of the main issues when comparing fusion data for different interacting systems and in particular when addressing whether cross sections are enhanced or hindered, is the various ways that fusion data from different systems are presented and compared. While various reduced units may be found in the literature \cite{Canto_2008,CANTO200951,SHORTO200977}, we have chosen to present our results using those defined by Gomes {\it et al.\/} \cite{redxsec} which  eliminate  the  so-called ``geometrical effects''. These reduced units, also referred as the ``simplified traditional method'', are completely model independent and appropriate when used for systems of similar masses  \cite{canto15}, such as those considered in this work.

In this representation:
\begin{equation}
E_\mathrm{red} = E_\mathrm{c.m.}\times (A^{1/3}_p + A^{1/3}_t) / (Z_{t}Z_{p}) \\
\end{equation}
and 
\begin{equation}
\sigma_\mathrm{red} = \sigma / (A^{1/3}_p + A^{1/3}_t)^2,
\end{equation}
where $E_\mathrm{c.m.}$ is the energy in the center-of-mass system in MeV, $\sigma$ is the measured cross section in mb, and $A_{p}$, $A_{t}$, $Z_{p}$, and $Z_{t}$ refer to the mass ($A$) and the nuclear charge ($Z$) of the projectile ($p$) and target ($t$) nuclei involved in the reaction. 

\begin{figure}
\centering
\includegraphics[width=\linewidth]{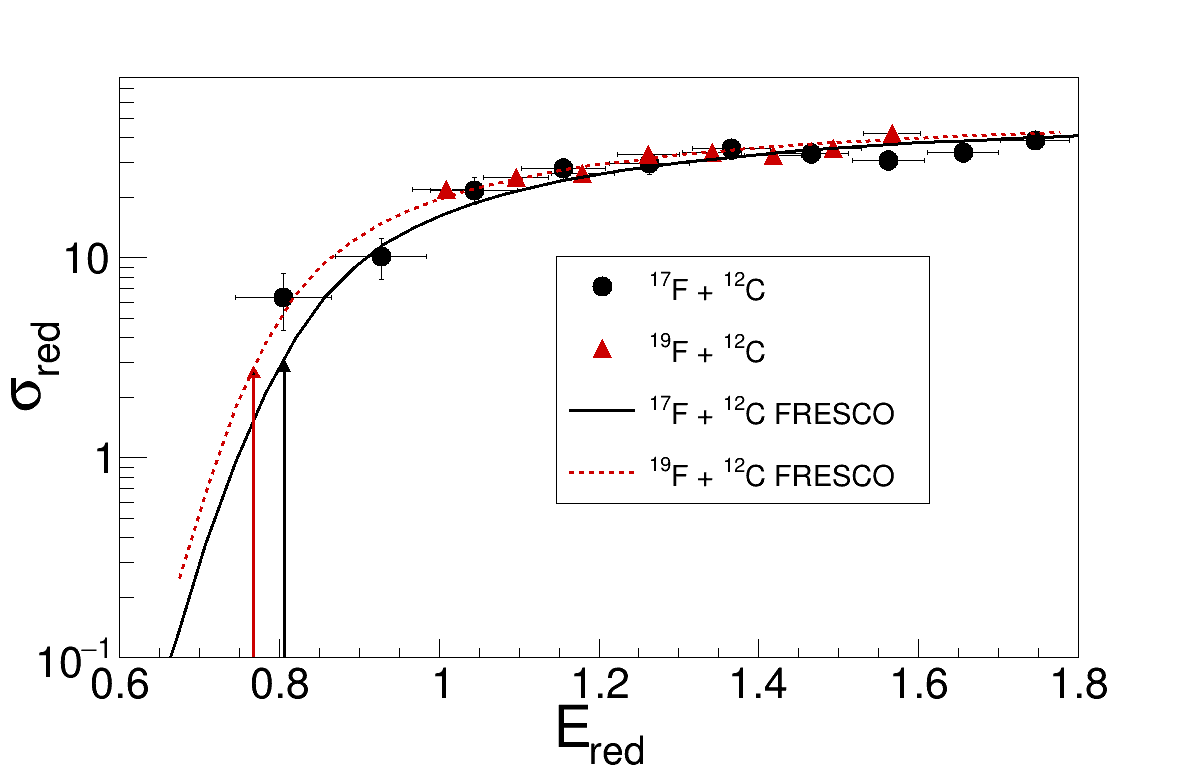}
\caption{Reduced total fusion cross sections for the $^{17}$F + $^{12}$C (black circles) and $^{19}$F + $^{12}$C (red triangles) systems measured in this work. The corresponding no coupling barrier penetration calculations using double folded real potentials are denoted by the solid and dotted lines, respectively. The fusion barrier heights extracted from the potentials are indicated by the vertical arrows.}
\label{fig:19F}
\end{figure}

\begin{figure}
\centering
\includegraphics[width=\linewidth]{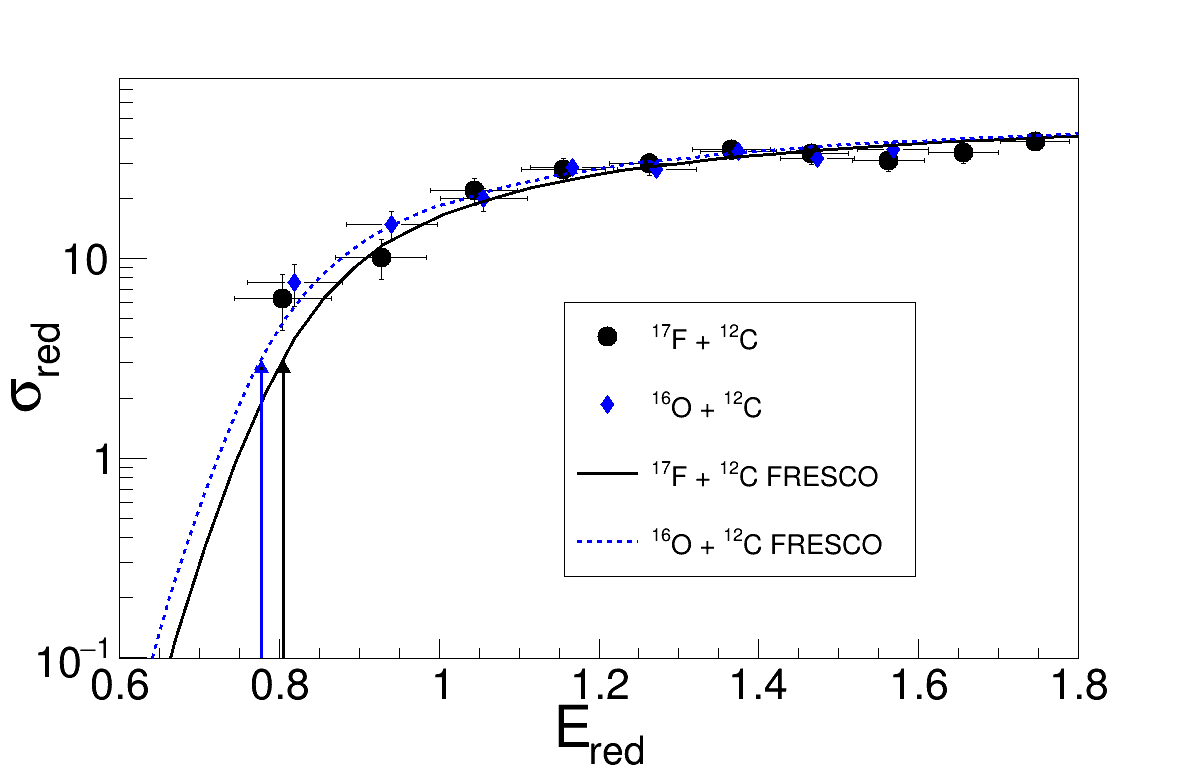}
\caption{Reduced total fusion cross sections for the $^{17}$F + $^{12}$C (black circles) and $^{16}$O + $^{12}$C (blue diamonds) systems measured in this work. The corresponding no coupling barrier penetration calculations using double folded real potentials are denoted by the solid and dotted lines, respectively. The fusion barrier heights extracted from the potentials are indicated by the vertical arrows.}
\label{fig:16O}
\end{figure}

In employing these reduced units we seek to minimize biases arising from ``trivial'' differences in Coulomb barrier heights and the $A^{1/3}$ nuclear radius variation which could ``wash out'' any structure effects that may be evident in the data \cite{redxsec} while at the same time retaining any ``static'' effects due to the increased size of the halo state. Using this convention, our measurements of the $^{17}$F + $^{12}$C fusion excitation function are plotted together with those for the $^{19}$F + $^{12}$C and $^{16}$O + $^{12}$C systems carried out under the same experimental conditions using the \textit{Encore} detector at FSU in Figs.\ \ref{fig:19F} and \ref{fig:16O}, respectively.

Also displayed in Figs.\ \ref{fig:19F} and \ref{fig:16O} are theoretical fusion excitation functions for the three systems calculated with the code {\sc fresco} \cite{Tho88}. The real parts of the nuclear potentials were obtained using the double folding procedure and the M3Y
nucleon-nucleon interaction \cite{Sat79}. The $^{12}$C, $^{16}$O, and $^{19}$F nuclear matter densities were derived from the experimental charge 
densities of Refs.\ \cite{Car80}, \cite{Lah82}, and \cite{Hal73}, respectively by unfolding the proton charge density and assuming that 
$\rho_\mathrm{Nuc} = (1 + N/Z)\rho_p$. The $^{17}$F nuclear matter density was taken from Ref.\ \cite{RIPL-3}. 
The double-folded potentials were calculated with the code {\sc dfpot} \cite{Coo82}. An imaginary potential of Woods-Saxon squared form with
parameters $W = 50$ MeV, $r = 1.0 \times (12^{1/3} + A_p^{1/3})$ fm, $a = 0.3$ fm effectively reproduced the incoming-wave boundary condition, the
total fusion cross section being calculated as the absorption by this potential. In all three systems it was found that no channel couplings were needed to reproduce the experimental results, which were well described by barrier penetration calculations \cite{Wong}. The fusion barriers ($V_\mathrm{b}$), fusion radii ($R_\mathrm{b}$), and barrier curvatures ($\hbar\omega$) extracted from the double folded potentials are given in Table \ref{table:table2}.

\begin{table}
\begin{center}
\begin{tabular}{ |c|c|c|c| } 
 \hline
System & $V_\mathrm{b}$ (MeV) & $R_\mathrm{b}$ (fm) & $\hbar\omega$ (MeV)   \\ 
 \hline
 
$^{16}$O + $^{12}$C & 7.76 & 8.01 & 2.67	\\
$^{19}$F + $^{12}$C & 8.37 & 8.36 & 2.56	\\
$^{17}$F + $^{12}$C & 8.95 & 8.03 & 3.04 \\
 \hline
\end{tabular}
\caption{Fusion barrier parameters for the systems studied in this work extracted from the double-folded potentials used in the {\sc fresco} calculations. When reduced according to the scheme of Gomes {\it et al.\/} \cite{redxsec} the barrier heights $V_\mathrm{b}$ for the three systems are almost identical: 0.78, 0.77, and 0.81} for $^{16}$O + $^{12}$C, $^{19}$F + $^{12}$C, and $^{17}$F + $^{12}$C, respectively.
\label{table:table2}
\end{center}
\end{table}

The measured total fusion cross sections for the $^{17}$F + $^{12}$C (black circles) and $^{19}$F + $^{12}$C (red triangles) systems plotted in Fig.\ \ref{fig:19F} agree with each other well over the measured energy range when compared in reduced units. No enhancement or reduction of the $^{17}$F fusion cross section compared with that for its stable counterpart is observed. The {\sc fresco} calculations for both systems are also shown in Fig.\ \ref{fig:19F} as solid and dotted lines for the $^{17}$F + $^{12}$C and $^{19}$F + $^{12}$C systems, respectively. The extracted barrier parameters for $^{17}$F and $^{19}$F are in close agreement, suggesting that the weak binding of $^{17}$F and the possible proton halo nature of its low-lying $1/2^+$ excited state have little or no influence on the fusion cross section in the measured energy range, in accord with previous results for the heavy $^{17}$F + $^{208}$Pb \cite{ErnstFlourine,flourinereaction} and medium mass $^{17}$F + $^{58}$Ni \cite{Yang2021} systems.

The total fusion cross sections for the $^{17}$F + $^{12}$C (black circles) and $^{16}$O + $^{12}$C (blue diamonds) systems are compared in Fig.\ \ref{fig:16O}. No enhancement {or reduction} of the $^{17}$F fusion cross section compared to that for its $^{16}$O core is observed when the excitation functions are plotted in reduced units.
The {\sc fresco} calculations for these two systems are also shown in Fig.\ \ref{fig:16O} as the solid and dotted lines, respectively. The $V_\mathrm{b}$ values extracted for $^{17}$F and $^{16}$O are significantly different (see Table \ref{table:table2}), as expected due to their differing $Z$ values. However, their reduced values, 0.81 and 0.78, respectively, are almost identical, suggesting that the valence proton has little or no influence on the $^{17}$F fusion cross section over the measured energy range. The $R_\mathrm{b}$ values are also similar: 8.03 fm and 8.01 fm, respectively, again suggesting that the valence proton has minimal influence on the fusion.

The lack of enhancement of the $^{17}$F total fusion cross sections compared to those for its $^{16}$O core is consistent with calculated values of the ground state r.m.s. matter radius of $^{17}$F which yield values similar to that of the ground state of $^{16}$O (see, e.g., Ref.\ \cite{Zha03}). While the r.m.s. radius of the $^{17}$F 0.495 MeV $1/2^+$ excited state is significantly larger, commensurate with its proposed halo status, the lack of enhancement of the $^{17}$F fusion cross section strongly suggests that its influence on the fusion process is small, either through coupling effects on the fusion barrier height or directly as a result of fusion of the $^{17}$F after being excited to this state. 

In order further to test this conclusion, coupled channel (CC) calculations were also performed with {\sc fresco}. The ``bare'', no 
coupling potential was the same as that used in the barrier penetration calculation described previously, the total fusion cross section in 
this case being calculated as the sum of the absorption by the Woods-Saxon squared imaginary potential in all channels.

Couplings to the 0.495-MeV $1/2^+$ state of $^{17}$F and the 4.44-MeV $2^+$ state of $^{12}$C were included using standard collective model
form factors. The $B(E2)$ for the $^{17}$F coupling was taken from Ref.\ \cite{Ber07} and the nuclear deformation length was derived from 
this value assuming the collective model and a radius of $1.3 \times 17^{1/3}$ fm. The $^{12}$C $B(E2)$ was taken from Ref.\ \cite{Ram01} 
and the nuclear deformation length from Ref.\ \cite{Kee15}. 

\begin{figure}
\includegraphics[clip=,width=\columnwidth]{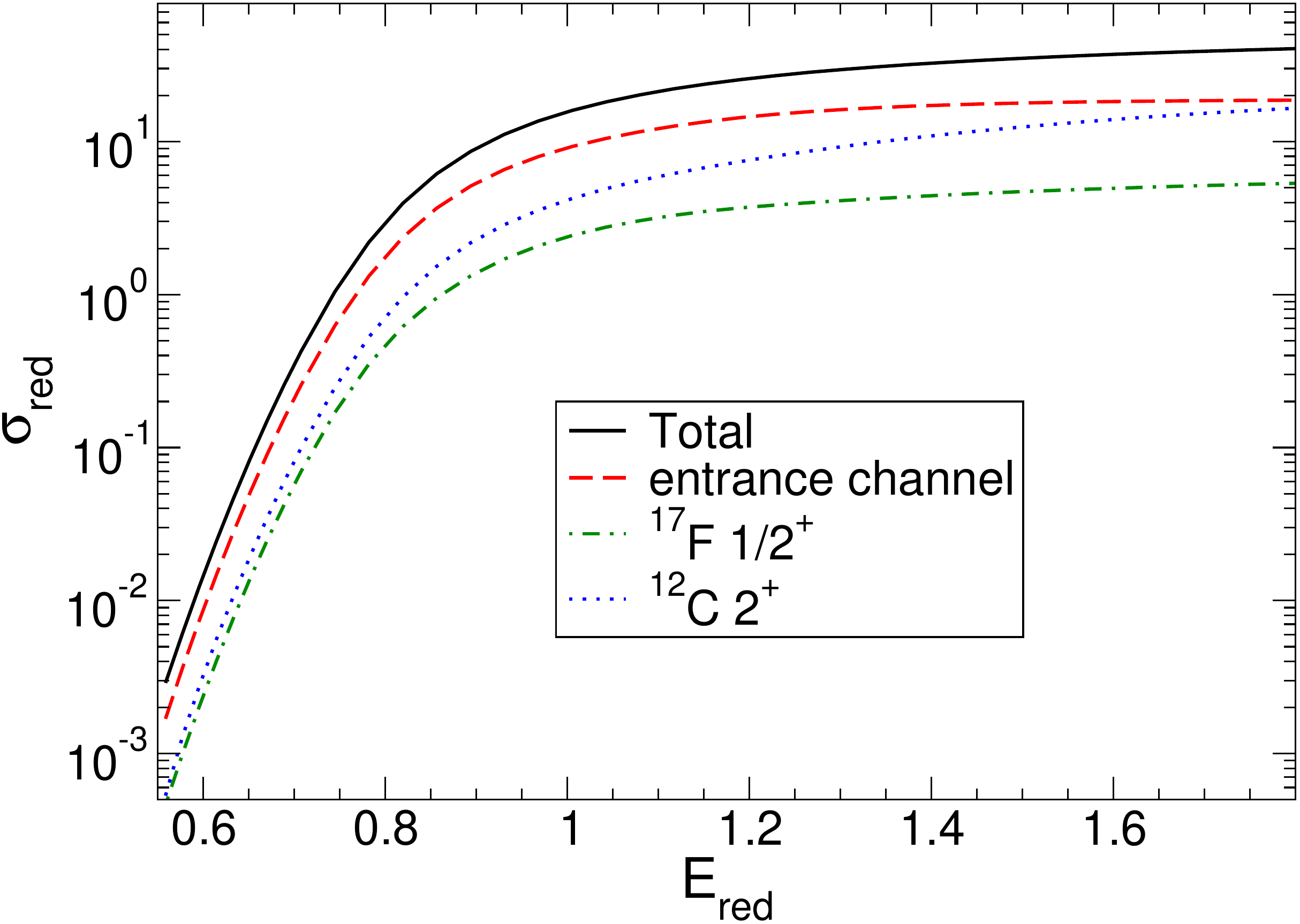}
\caption{Fusion excitation functions for $^{17}$F + $^{12}$C calculated using the code {\sc fresco}. The solid curve denotes the 
total fusion cross section, the dashed curve fusion for the entrance channel (projectile and target in their respective ground states), the
dot-dashed curve fusion for the channel with the $^{17}$F in its 0.495-MeV $1/2^+$ excited state and the $^{12}$C in its ground state, and
the dotted curve fusion for the channel with the $^{17}$F in its ground state and the $^{12}$C in its 4.44-MeV $2^+$ excited state.}
\label{fig6}
\end{figure}

It is possible to calculate the absorption by the imaginary potential for individual 
channels using {\sc fresco}, equivalent to the fusion cross section in the model used here. In Fig.\ \ref{fig6} we present the excitation
functions for the total fusion and for each of the following channels: the entrance channel with both the $^{17}$F projectile and the $^{12}$C
target in their respective ground states, the $^{17}$F in its 0.495-MeV $1/2^+$ excited state and the $^{12}$C in its ground state, and
the $^{17}$F in its ground state and the $^{12}$C in its 4.44-MeV $2^+$ excited state. Mutual excitation was not considered.
The data are omitted for the sake of clarity, but they are well reproduced by the calculated total fusion excitation function.

The fusion excitation function obtained from the no-coupling calculation (not shown) is visually indistinguishable from the CC result on 
the scale of the figure down to values of $E_\mathrm{red} = 0.75$, a slight enhancement due to the coupling being
visible at lower energies. Careful comparison of the two calculations indicates that the coupling induces enhancement of the
total fusion cross section below the Coulomb barrier and suppression above it, although the latter in particular is too small to be seen on
a logarithmic plot. The coupling effects of both the 0.495-MeV $1/2^+$ $^{17}$F excited state and the 4.44-MeV $2^+$ $^{12}$C
excited state on the fusion may thus be considered negligible over the measured energy range, confirming the conclusion of the barrier penetration calculation. In addition, the breakdown of the 
total fusion cross section into its individual channels shows that the contribution from fusion with the $^{17}$F in its 0.495-MeV $1/2^+$
excited state is small over the measured $E_\mathrm{red}$ range, being about an order of magnitude smaller than the total. By contrast,
fusion with the $^{12}$C in its 4.44-MeV $2^+$ excited state is much more important, becoming comparable to fusion from the entrance
channel at the highest $E_\mathrm{red}$ values. The calculations predict that fusion with the $^{17}$F in its 0.495-MeV $1/2^+$ excited
state becomes more important at sub-barrier energies, but the trend suggests that it would only make a significant contribution to the
total fusion in the relatively deep sub-barrier region where the total fusion cross section is small, only becoming equal to fusion with
the $^{12}$C in its 4.44-MeV $2^+$ excited state at $E_\mathrm{red} \approx 5.5$, where fusion from the entrance channel still dominates the
total. 

The calculations presented in Fig.\ \ref{fig6} used the same double-folded real potential in all channels, calculated using the $^{17}$F
ground state density of Ref.\ \cite{RIPL-3}. Thus they do not take into account the extended size of the 0.495-MeV $1/2^+$ proton halo
state of $^{17}$F. However, test calculations where the diagonal potential for the channel with the $^{17}$F in this state was re-calculated
using the appropriate $^{17}$F matter density of Ref.\ \cite{Zha03} which explicitly includes the proton halo gave almost identical
results.

\section{Discussion and Conclusions}

In summary, an experimental campaign to study the influence of the structure of the weakly-bound, proton drip-line nucleus $^{17}$F on its total fusion cross section with $^{12}$C was carried out using the \emph{Encore} active-target detector system at FSU. \emph{Encore} measures energy losses as the beam travels through the detector using a segmented anode. The total fusion excitation function over a wide energy range can be measured with absolute normalization and without changing the beam energy. Systematic measurements with its stable counterparts $^{16}$O and $^{19}$F were also performed under the same experimental conditions. 

The comparison of the data in reduced units presented in this work indicates no special influence on the total fusion cross section over the measured energy range due to the specific structure of $^{17}$F, in particular the proton halo nature of its its low-lying first excited state, confirming previous findings for the heavy $^{17}$F + $^{208}$Pb \cite{ErnstFlourine} and medium mass $^{17}$F + $^{58}$Ni \cite{Yang2021} systems. The lack of a fusion enhancement in the present work with the low Z target nucleus $^{12}$C provides a further test of the proposal by Xue-Ying {\it et al.\/} \cite{shielding} that in the $^{17}$F + $^{208}$Pb system no enhancement in the fusion cross section was observed because the large Coulomb field of the $^{208}$Pb target polarizes the $^{17}$F such that the proton follows the $^{16}$O core in its interaction with the target. Together with the results for the $^{17}$F + $^{58}$Ni \cite{Yang2021} system the present work suggests that, at least for incident energies down to the top of the Coulomb barrier, a polarization effect of this type cannot account for the lack of enhancement in the total fusion cross section for $^{17}$F since the charge products of the systems analyzed now extend from 738 through 252 down to 54 and no enhancement is observed. 

Coupled channel calculations showed that the influence of coupling to the $^{17}$F proton halo state on the
total fusion cross section for the $^{17}$F + $^{12}$C system was suppression above the barrier and enhancement 
below (with respect to the no-coupling calculation) although the size of the effect was negligible except for
energies $E_\mathrm{red} < 0.7$. The CC calculations further indicated that fusion from the 
channel with $^{17}$F propagating in the halo state makes only a small contribution to the total fusion. Thus, the halo
nature of the 0.495-MeV $1/2^+$ excited state of $^{17}$F is seen to have little or no influence on the total fusion
in the energy range studied here, explaining the similarity of the fusion cross sections for the $^{17}$F, $^{19}$F,
and $^{16}$O + $^{12}$C systems when plotted in reduced units. 

Analogous CC calculations performed for the $^{17}$F + $^{58}$Ni and $^{208}$Pb 
systems showed qualitatively similar behavior in that the coupling to the 0.495-MeV $1/2^+$ excited state of $^{17}$F induced enhancement of
the total fusion cross section below the Coulomb barrier and suppression above, albeit the coupling effects were more important and increased with
increasing target charge, a well attested effect. The coupling effect is consistent with that seen in the CDCC calculations for the
$^{17}$F + $^{58}$Ni system \cite{Yang2021}. The breakdown of the total fusion cross section by channel showed that the contribution from
the channel with the $^{17}$F propagating in its 0.495-MeV $1/2^+$ excited state was again small for both systems. 

Taking the results for the $^{17}$F + $^{12}$C, $^{17}$F + $^{58}$Ni, and $^{17}$F + $^{208}$Pb systems all together  
we see a consistent picture emerging. By far the most important influence of the $^{17}$F 0.495-MeV $1/2^+$ 
excited state on the total fusion is via its coupling effect on the fusion barrier, and the size of this effect scales with increasing target charge, as for
any inelastic coupling. The main influence of the halo nature of this state on the total fusion cross section will thus only be indirectly via
any impact it may have on the coupling strength linking it to the ground state. Our final conclusion is, therefore, 
that as regards the total fusion, $^{17}$F essentially behaves in a similar way to any other nucleus with strongly coupled excited states, 
neither its weak binding nor the halo nature of its first excited state making any striking impact on this observable.

The self-normalizing and efficient capabilities of the \textit{Encore} detector make it an ideal system systematically to study the fusion cross sections for other proposed proton-halo nuclei: $^{8}$B \cite{BorNi, BorSi}, $^{17}$Ne \cite{neon17}, and $^{27}$P \cite{phosphorus27} and thus determine whether there are any specific structure related effect on the total fusion cross section for these nuclei at energies near the top of the Coulomb barrier. Other ``detector targets'' such as Ne, Ar, Kr and Xe could be used further to explore the Z dependence of the fusion cross sections for exotic proton-halo nuclei.

\section*{Acknowledgements}

This work was supported by the National Science Foundation under grants PHY-1712953 and PHY-2012522, and by the State of Florida.

\clearpage
\bibliographystyle{unsrt}
\bibliography{main.bbl}

\begin{thebibliography}{10}

\bibitem{BackReview}
B.~B. Back, H.~Esbensen, C.~L. Jiang, and K.~E. Rehm.
\newblock Recent developments in heavy-ion fusion reactions.
\newblock {\em Rev. Mod. Phys.}, 86:317--360, 2014.

\bibitem{jiang12C}
C.~L. Jiang, D.~Santiago-Gonzalez, S.~Almaraz-Calderon, K.~E. Rehm, B.~B. Back,
  K.~Auranen, M.~L. Avila, A.~D. Ayangeakaa, S.~Bottoni, M.~P. Carpenter,
  C.~Dickerson, B.~DiGiovine, J.~P. Greene, C.~R. Hoffman, R.~V.~F. Janssens,
  B.~P. Kay, S.~A. Kuvin, T.~Lauritsen, R.~C. Pardo, J.~Sethi, D.~Seweryniak,
  R.~Talwar, C.~Ugalde, S.~Zhu, D.~Bourgin, S.~Courtin, F.~Haas, M.~Heine,
  G.~Fruet, D.~Montanari, D.~G. Jenkins, L.~Morris, A.~Lefebvre-Schuhl,
  M.~Alcorta, X.~Fang, X.~D. Tang, B.~Bucher, C.~M. Deibel, and S.~T. Marley.
\newblock Reaction rate for carbon burning in massive stars.
\newblock {\em Phys. Rev. C}, 97:012801, 2018.

\bibitem{CarnelliPrl}
P.~F.~F. Carnelli, S.~Almaraz-Calderon, K.~E. Rehm, M.~Albers, M.~Alcorta,
  P.~F. Bertone, B.~Digiovine, H.~Esbensen, J.~O.~Fern\'andez Niello,
  D.~Henderson, C.~L. Jiang, J.~Lai, S.~T. Marley, O.~Nusair, T.~Palchan-Hazan,
  R.~C. Pardo, M.~Paul, and C.~Ugalde.
\newblock Measurements of fusion reactions of low-intensity radioactive carbon
  beams on $^{12}\mathrm{C}$ and their implications for the understanding of
  x-ray bursts.
\newblock {\em Phys. Rev. Lett.}, 112:192701, 2014.

\bibitem{keeley07}
N.~Keeley, R.~Raabe, N.~Alamanos, and J.L. Sida.
\newblock Fusion and direct reactions of halo nuclei at energies around the
  coulomb barrier.
\newblock {\em Progress in Particle and Nuclear Physics}, 59(2):579 -- 630,
  2007.

\bibitem{tanihata}
I.~Tanihata, H.~Hamagaki, O.~Hashimoto, Y.~Shida, N.~Yoshikawa, K.~Sugimoto,
  O.~Yamakawa, T.~Kobayashi, and N.~Takahashi.
\newblock Measurements of interaction cross sections and nuclear radii in the
  light $p$-shell region.
\newblock {\em Phys. Rev. Lett.}, 55:2676--2679, 1985.

\bibitem{tanihatahe}
I.~Tanihata, H.~Hamagaki, O.~Hashimoto, S.~Nagamiya, Y.~Shida, N.~Yoshikawa,
  O.~Yamakawa, K.~Sugimoto, T.~Kobayashi, D.E. Greiner, N.~Takahashi, and
  Y.~Nojiri.
\newblock Measurements of interaction cross sections and radii of he isotopes.
\newblock {\em Phys. Lett. B}, 160(6):380 -- 384, 1985.

\bibitem{canto15}
L.F. Canto, P.R.S. Gomes, R.~Donangelo, J.~Lubian, and M.S. Hussein.
\newblock Recent developments in fusion and direct reactions with weakly bound
  nuclei.
\newblock {\em Physics Reports}, 596:1 -- 86, 2015.

\bibitem{balantekin98}
A.~B. Balantekin and N.~Takigawa.
\newblock Quantum tunneling in nuclear fusion.
\newblock {\em Rev. Mod. Phys.}, 70:77--100, 1998.

\bibitem{ErnstFlourine}
K.~E. Rehm, H.~Esbensen, C.~L. Jiang, B.~B. Back, F.~Borasi, B.~Harss, R.~V.~F.
  Janssens, V.~Nanal, J.~Nolen, R.~C. Pardo, M.~Paul, P.~Reiter, R.~E. Segel,
  A.~Sonzogni, J.~Uusitalo, and A.~H. Wuosmaa.
\newblock Fusion cross sections for the proton drip line nucleus $^{17}{F}$ at
  energies below the coulomb barrier.
\newblock {\em Phys. Rev. Lett.}, 81:3341--3344, 1998.

\bibitem{kolatareview}
J.~J. Kolata, V.~Guimarães, and E.~F. Aguilera.
\newblock Elastic scattering, fusion, and breakup of light exotic nuclei.
\newblock {\em The European Physical Journal A}, 52(123), 2016.

\bibitem{6He}
V.~Scuderi, A.~Di~Pietro, P.~Figuera, M.~Fisichella, F.~Amorini, C.~Angulo,
  G.~Cardella, E.~Casarejos, M.~Lattuada, M.~Milin, A.~Musumarra, M.~Papa,
  M.~G. Pellegriti, R.~Raabe, F.~Rizzo, N.~Skukan, D.~Torresi, and M.~Zadro.
\newblock Fusion and direct reactions for the system $^{6}${He} $+$ $^{64}${Zn}
  at and below the coulomb barrier.
\newblock {\em Phys. Rev. C}, 84:064604, 2011.

\bibitem{8He}
{Parkar, V.V.}, {Marquinez, G.}, {Martel, I.}, {S\'anchez-Ben\'{\i}tez, A.M.},
  {Acosta, L.}, {Berjillos, R.}, {Due\~nas, J.}, {Flores, J.L.}, {Bol\'{\i}var,
  J.P.}, {Padilla, A.}, {Alvarez, M.A.G.}, {Beaumel, D.}, {Borge, M.J.G.},
  {Chbihi, A.}, {Cruz, C.}, {Cubero, M.}, {Fernandez Garcia, J.P.},
  {Fern\'andez Mart\'{\i}nez, B.}, {Gomez Camacho, J.}, {Keeley, N.},
  {Labrador, J.A.}, {Marquis, M.}, {Mazzocco, M.}, {Pakou, A.}, {Patronis, N.},
  {Pesudo, V.}, {Pierroutsakou, D.}, {Raabe, R.}, {Rusek, K.}, {Silvestri, R.},
  {Standylo, L.}, {Strojek, I.}, {Soic, N.}, {Tengblad, O.}, {Wolski, R.}, and
  {Ziad, A.H.}
\newblock Fusion of 8he with 206pb around coulomb barrier energies.
\newblock {\em EPJ Web of Conferences}, 17:16009, 2011.

\bibitem{11Li}
A.~M. Vinodkumar, W.~Loveland, R.~Yanez, M.~Leonard, L.~Yao, P.~Bricault,
  M.~Dombsky, P.~Kunz, J.~Lassen, A.~C. Morton, D.~Ottewell, D.~Preddy, and
  M.~Trinczek.
\newblock Interaction of ${}^{11}${Li} with ${}^{208}${Pb}.
\newblock {\em Phys. Rev. C}, 87:044603, 2013.

\bibitem{11Be}
C.~Signorini, A.~Yoshida, Y.~Watanabe, D.~Pierroutsakou, L.~Stroe, T.~Fukuda,
  M.~Mazzocco, N.~Fukuda, Y.~Mizoi, M.~Ishihara, H.~Sakurai, A.~Diaz-Torres,
  and K.~Hagino.
\newblock Subbarrier fusion in the systems $^{11,10}${Be}+$^{209}${Bi}.
\newblock {\em Nuc. Phys. A}, 735(3):329 -- 344, 2004.

\bibitem{11Beanalysis}
D.~J. Hinde and M.~Dasgupta.
\newblock Systematic analysis of above-barrier fusion of
  $^{9,10,11}\mathrm{Be}$$+$$^{209}\mathrm{Bi}$.
\newblock {\em Phys. Rev. C}, 81:064611, 2010.

\bibitem{halostuff}
P~G Hansen, A~S Jensen, and B~Jonson.
\newblock Nuclear halos.
\newblock {\em Annual Review of Nuclear and Particle Science}, 45(1):591--634,
  1995.

\bibitem{BorNi}
E.~F. Aguilera, P.~Amador-Valenzuela, E.~Martinez-Quiroz, D.~Lizcano,
  P.~Rosales, H.~Garc\'{\i}a-Mart\'{\i}nez, A.~G\'omez-Camacho, J.~J. Kolata,
  A.~Roberts, L.~O. Lamm, G.~Rogachev, V.~Guimar\~aes, F.~D. Becchetti,
  A.~Villano, M.~Ojaruega, M.~Febbraro, Y.~Chen, H.~Jiang, P.~A. DeYoung, G.~F.
  Peaslee, C.~Guess, U.~Khadka, J.~Brown, J.~D. Hinnefeld, L.~Acosta,
  E.~S.~Rossi Jr, J.~F.~P. Huiza, and T.~L. Belyaeva.
\newblock Near-barrier fusion of the $^{8}${B}+$^{58}${Ni} proton-halo system.
\newblock {\em Phys. Rev. Lett.}, 107:092701, 2011.

\bibitem{BorSi}
A.~Pakou, E.~Stiliaris, D.~Pierroutsakou, N.~Alamanos, A.~Boiano, C.~Boiano,
  D.~Filipescu, T.~Glodariu, J.~Grebosz, A.~Guglielmetti, M.~La~Commara,
  M.~Mazzocco, C.~Parascandolo, K.~Rusek, A.~M. S\'anchez-Ben\'{\i}tez,
  C.~Signorini, O.~Sgouros, F.~Soramel, V.~Soukeras, E.~Strano, L.~Stroe,
  N.~Toniolo, D.~Torresi, and K.~Zerva.
\newblock Fusion cross sections of ${}^{8}${B} + ${}^{28}${Si} at near-barrier
  energies.
\newblock {\em Phys. Rev. C}, 87:014619, 2013.

\bibitem{17FHalo}
R.~Morlock, R.~Kunz, A.~Mayer, M.~Jaeger, A.~M\"uller, J.~W. Hammer, P.~Mohr,
  H.~Oberhummer, G.~Staudt, and V.~K\"olle.
\newblock Halo properties of the first $1/{2}^{+}$ state in $^{17}${F} from the
  $^{16}${O}(p,$\gamma$)$^{17}${F} reaction.
\newblock {\em Phys. Rev. Lett.}, 79:3837--3840, 1997.

\bibitem{FlNi}
J.F. Liang et~al.
\newblock Dynamic polarization in the coulomb breakup of loosely bound
  $^{17}${F}.
\newblock {\em Phys. Lett. B}, 681:22--25, 2009.

\bibitem{FlPb}
Wang Qi, Han Jian-Long, Xiao Zhi-Gang, Xu~Hu-Shan, Sun Zhi-Yu, Hu~Zheng-Guo,
  Zhang Xue-Ying, Wang Hong-Wei, Mao Rui-Shi, Yuan Xiao-Hua, Xu~Zhi-Guo, Zhao
  Tie-Cheng, Zhang Hong-Bin, Xu~Hua-Gen, Qi~Hui-Rong, Wang Yue, Jia Fei,
  Wu~Li-Jie, Ding Xian-Li, Gao Qi, Gao Hui, Li~Song-Lin, Bai Zhen, Xiao
  Guo-Qing, Jin Gen-Ming, Ren Zhong-Zhou, Zhou Shan-Gui, and Sergey Yu-Kun.
\newblock Exotic behaviour of angular dispersion of weakly bound nucleus
  $^{17}${F} at small angles.
\newblock {\em Chinese Physics Letters}, 23(7):1731--1733, 2006.

\bibitem{flourinereaction}
Zhang H.Q. et~al. Zhang~G.L., Zhang~C.L.
\newblock Quasi-elastic scattering of the proton drip line nucleus $^{17}${F}
  on $^{12}${C} at 60 mev.
\newblock {\em Eur. Phys. J. A}, 48, 2012.

\bibitem{Mazz2010}
M.~Mazzocco, C.~Signorini, D.~Pierroutsakou, T.~Glodariu, A.~Boiano, C.~Boiano,
  F.~Farinon, P.~Figuera, D.~Filipescu, L.~Fortunato, A.~Guglielmetti,
  G.~Inglima, M.~La~Commara, M.~Lattuada, P.~Lotti, C.~Mazzocchi, P.~Molini,
  A.~Musumarra, A.~Pakou, C.~Parascandolo, N.~Patronis, M.~Romoli, M.~Sandoli,
  V.~Scuderi, F.~Soramel, L.~Stroe, D.~Torresi, E.~Vardaci, and A.~Vitturi.
\newblock Reaction dynamics for the system $^{17}\mathrm{F}+^{58}\mathrm{Ni}$
  at near-barrier energies.
\newblock {\em Phys. Rev. C}, 82:054604, Nov 2010.

\bibitem{Liang2000}
J.F. Liang, J.R. Beene, H.~Esbensen, A.~Galindo-Uribarri, J.~{Gomez del Campo},
  C.J. Gross, M.L. Halbert, P.E. Mueller, D.~Shapira, D.W. Stracener, and R.L.
  Varner.
\newblock Breakup of weakly bound $^{17}$f well above the coulomb barrier.
\newblock {\em Physics Letters B}, 491(1):23 -- 28, 2000.

\bibitem{Lian2003}
J.~F. Liang, J.~R. Beene, A.~Galindo-Uribarri, J.~Gomez~del Campo, C.~J. Gross,
  P.~A. Hausladen, P.~E. Mueller, D.~Shapira, D.~W. Stracener, R.~L. Varner,
  J.~D. Bierman, H.~Esbensen, and Y.~Larochelle.
\newblock Breakup of ${}^{17}\mathrm{F}$ on ${}^{208}\mathrm{Pb}$ near the
  coulomb barrier.
\newblock {\em Phys. Rev. C}, 67:044603, Apr 2003.

\bibitem{Sig2010}
{Signorini, C.}, {Pierroutsakou, D.}, {Martin, B.}, {Mazzocco, M.}, {Glodariu,
  T.}, {Bonetti, R.}, {Guglielmetti, A.}, {La Commara, M.}, {Romoli, M.},
  {Sandoli, M.}, {Vardaci, E.}, {Esbensen, H.}, {Farinon, F.}, {Molini, P.},
  {Parascandolo, C.}, {Soramel, F.}, {Sidortchuk, S.}, and {Stroe, L.}
\newblock Interaction of $^{17}${F} with a $^{208}${Pb} target below the
  coulomb barrier.
\newblock {\em Eur. Phys. J. A}, 44(1):63--69, 2010.

\bibitem{Romoli2004}
M.~Romoli, E.~Vardaci, M.~Di~Pietro, A.~De~Francesco, A.~De~Rosa, G.~Inglima,
  M.~La~Commara, B.~Martin, D.~Pierroutsakou, M.~Sandoli, M.~Mazzocco,
  T.~Glodariu, P.~Scopel, C.~Signorini, R.~Bonetti, A.~Guglielmetti,
  F.~Soramel, L.~Stroe, J.~Greene, A.~Heinz, D.~Henderson, C.~L. Jiang, E.~F.
  Moore, R.~C. Pardo, K.~E. Rehm, A.~Wuosmaa, and J.~F. Liang.
\newblock Measurements of $^{17}\mathrm{F}$ scattering by $^{208}\mathrm{Pb}$
  with a new type of large solid angle detector array.
\newblock {\em Phys. Rev. C}, 69:064614, Jun 2004.

\bibitem{Yang2021}
L.~Yang, C.J. Lin, H.~Yamaguchi, Jin Lei, P.W. Wen, M.~Mazzocco, N.R. Ma, L.J.
  Sun, D.X. Wang, G.X. Zhang, K.~Abe, S.M. Cha, K.Y. Chae, A.~Diaz-Torres, J.L.
  Ferreira, S.~Hayakawa, H.M. Jia, D.~Kahl, A.~Kim, M.S. Kwag, M.~{La Commara},
  R.~{Navarro PÃ©rez}, C.~Parascandolo, D.~Pierroutsakou, J.~Rangel,
  Y.~Sakaguchi, C.~Signorini, E.~Strano, X.X. Xu, F.~Yang, Y.Y. Yang, G.L.
  Zhang, F.P. Zhong, and J.~Lubian.
\newblock Insight into the reaction dynamics of proton drip-line nuclear system
  $^{17}${F}+$^{58}${Ni} at near-barrier energies.
\newblock {\em Physics Letters B}, 813:136045, 2021.

\bibitem{shielding}
Xue ying He, Qin Dong, and Li~Ou.
\newblock Shielding effects in fusion reactions with a proton-halo nucleus.
\newblock {\em Chinese Physics C}, 44(5):054108, 2020.

\bibitem{resolut}
I.~{Wiedenh{\"o}ver}, L.~T. {Baby}, D.~{Santiago-Gonzalez}, A.~{Rojas}, J.~C.
  {Blackmon}, G.~V. {Rogachev}, J.~{Belarge}, E.~{Koshchiy}, A.~N. {Kuchera},
  L.~E. {Linhardt}, J.~{Lail}, K.~T. {Macon}, M.~{Matos}, and B.~C. {Rascol}.
\newblock {Studies of Exotic Nuclei at the Resolut Facility of Florida State
  University}.
\newblock In {\em Fission and Properties of Neutron-Rich Nuclei - Proceedings
  of the Fifth International Conference on ICFN5. Edited by Hamilton Joseph H
  \& Ramayya Akunuri V. Published by World Scientific Publishing Co. Pte. Ltd},
  pages 144--151, September 2014.

\bibitem{CarnelliNim}
P.F.F. Carnelli, S.~Almaraz-Calderon, K.E. Rehm, M.~Albers, M.~Alcorta, P.F.
  Bertone, B.~Digiovine, H.~Esbensen, J.~Fernández Niello, D.~Henderson, C.L.
  Jiang, J.~Lai, S.T. Marley, O.~Nusair, T.~Palchan-Hazan, R.C. Pardo, M.~Paul,
  and C.~Ugalde.
\newblock Multi-sampling ionization chamber (music) for measurements of fusion
  reactions with radioactive beams.
\newblock {\em Nuclear Instruments and Methods in Physics Research Section A:
  Accelerators, Spectrometers, Detectors and Associated Equipment}, 799:197 --
  202, 2015.

\bibitem{EncoreNim}
B.~W.~Asher et~al.
\newblock {\em Nuclear Instruments and Methods in Physics Research Section A:
  Accelerators, Spectrometers, Detectors and Associated Equipment}, to be
  published.

\bibitem{AvilaPrc}
M.~L. Avila, K.~E. Rehm, S.~Almaraz-Calderon, A.~D. Ayangeakaa, C.~Dickerson,
  C.~R. Hoffman, C.~L. Jiang, B.~P. Kay, J.~Lai, O.~Nusair, R.~C. Pardo,
  D.~Santiago-Gonzalez, R.~Talwar, and C.~Ugalde.
\newblock Experimental study of the astrophysically important
  $^{23}\mathrm{Na}(\ensuremath{\alpha},p)^{26}\mathrm{Mg}$ and
  $^{23}\mathrm{Na}(\ensuremath{\alpha},n)^{26}\mathrm{Al}$ reactions.
\newblock {\em Phys. Rev. C}, 94:065804, Dec 2016.

\bibitem{rashi}
R.~Talwar, M.~J. Bojazi, P.~Mohr, K.~Auranen, M.~L. Avila, A.~D. Ayangeakaa,
  J.~Harker, C.~R. Hoffman, C.~L. Jiang, S.~A. Kuvin, B.~S. Meyer, K.~E. Rehm,
  D.~Santiago-Gonzalez, J.~Sethi, C.~Ugalde, and J.~R. Winkelbauer.
\newblock Experimental study of $^{38}\mathrm{Ar}+\ensuremath{\alpha}$ reaction
  cross sections relevant to the $^{41}\mathrm{Ca}$ abundance in the solar
  system.
\newblock {\em Phys. Rev. C}, 97:055801, 2018.

\bibitem{lise}
O.B. Tarasov and D.~Bazin.
\newblock Lise++: Radioactive beam production with in-flight separators.
\newblock {\em Nuclear Instruments and Methods in Physics Research Section B:
  Beam Interactions with Materials and Atoms}, 266(19):4657 -- 4664, 2008.
\newblock Proceedings of the XVth International Conference on Electromagnetic
  Isotope Separators and Techniques Related to their Applications.

\bibitem{kovar16o}
D.~G. Kovar, D.~F. Geesaman, T.~H. Braid, Y.~Eisen, W.~Henning, T.~R. Ophel,
  M.~Paul, K.~E. Rehm, S.~J. Sanders, P.~Sperr, J.~P. Schiffer, S.~L. Tabor,
  S.~Vigdor, B.~Zeidman, and F.~W. Prosser.
\newblock Systematics of carbon- and oxygen-induced fusion on nuclei with
  $12\ensuremath{\le}a\ensuremath{\le}19$.
\newblock {\em Phys. Rev. C}, 20:1305--1331, 1979.

\bibitem{kolata16o}
J.J. Kolata, R.M. Freeman, F.~Haas, B.~Heusch, and A.~Gallmann.
\newblock On the fusion cross section for $^{16}${O} + $^{12}${C}.
\newblock {\em Phys. Lett. B}, 65(4):333 -- 336, 1976.

\bibitem{frohlich16o}
H.~Fröhlich, P.~Dück, W.~Galster, W.~Treu, H.~Voit, H.~Witt, W.~Kühn, and
  S.M. Lee.
\newblock Oscillations in the excitation function for complete fusion of
  $^{16}${O}+$^{12}${C} at low energies.
\newblock {\em Phys. Lett. B}, 64(4):408 -- 410, 1976.

\bibitem{chan16o}
Y.-D. Chan, H.~Bohn, R.~Vandenbosch, K.G. Bernhardt, J.G. Cramer, R.~Sielemann,
  and L.~Green.
\newblock Gross structure in $\gamma$-ray yields following the $^{16}${O}
  $^{12}${C} reaction.
\newblock {\em Nuc. Phys. A}, 303(3):500 -- 520, 1978.

\bibitem{Sperr16o}
P.~Sperr, S.~Vigdor, Y.~Eisen, W.~Henning, D.~G. Kovar, T.~R. Ophel, and
  B.~Zeidman.
\newblock Oscillations in the excitation function for complete fusion of
  $^{16}\mathrm{O}$+$^{12}\mathrm{C}$.
\newblock {\em Phys. Rev. Lett.}, 36:405--408, Feb 1976.

\bibitem{Das16o}
Binay Dasmahapatra, Bibiana Čujec, and Fouad Lahlou.
\newblock Fusion cross sections for $^{16}${O} + $^{13}${C} at low energies.
\newblock {\em Nuclear Physics A}, 394(1):301 -- 311, 1983.

\bibitem{Fraw16o}
A.~D. Frawley, N.~R. Fletcher, and L.~C. Dennis.
\newblock Resonances in the $^{16}\mathrm{O}$ + $^{12}\mathrm{C}$ fusion cross
  section between ${E}_{\mathrm{c}.\mathrm{m}.}=12 \mathrm{and} 20$ mev.
\newblock {\em Phys. Rev. C}, 25:860--865, Feb 1982.

\bibitem{Chri16o}
P.R. Christensen, Z.E. Switkowskiw, and R.A. Dayras.
\newblock Sub-barrier fusion measurements for $^{12}${C}+$^{16}${O}.
\newblock {\em Nuclear Physics A}, 280(1):189 -- 204, 1977.

\bibitem{Eyal16o}
Y.~Eyal, M.~Beckerman, R.~Chechik, Z.~Fraenkel, and H.~Stocker.
\newblock Nuclear size and boundary effects on the fusion barrier of oxygen
  with carbon.
\newblock {\em Phys. Rev. C}, 13:1527--1535, Apr 1976.

\bibitem{Kovar}
D.~G. Kovar, D.~F. Geesaman, T.~H. Braid, Y.~Eisen, W.~Henning, T.~R. Ophel,
  M.~Paul, K.~E. Rehm, S.~J. Sanders, P.~Sperr, J.~P. Schiffer, S.~L. Tabor,
  S.~Vigdor, B.~Zeidman, and F.~W. Prosser.
\newblock Systematics of carbon- and oxygen-induced fusion on nuclei with
  $12\ensuremath{\le}a\ensuremath{\le}19$.
\newblock {\em Phys. Rev. C}, 20:1305--1331, Oct 1979.

\bibitem{Anjos}
R.~M. Anjos, V.~Guimares, N.~Added, N.~Carlin~Filho, M.~M. Coimbra, L.~Fante,
  M.~C.~S. Figueira, E.~M. Szanto, C.~F. Tenreiro, and A.~Szanto~de Toledo.
\newblock Effect of the entrance channel mass asymmetry on the limitation of
  light heavy-ion fusion cross sections.
\newblock {\em Phys. Rev. C}, 42:354--362, Jul 1990.

\bibitem{Sperr19F}
P.~Sperr, T.~H. Braid, Y.~Eisen, D.~G. Kovar, F.~W. Prosser, J.~P. Schiffer,
  S.~L. Tabor, and S.~Vigdor.
\newblock Fusion cross sections of light heavy-ion systems: Resonances and
  shell effects.
\newblock {\em Phys. Rev. Lett.}, 37:321--323, Aug 1976.

\bibitem{Canto_2008}
L~F Canto, P~R~S Gomes, J~Lubian, L~C Chamon, and E~Crema.
\newblock Disentangling static and dynamic effects of low breakup threshold in
  fusion reactions.
\newblock {\em Journal of Physics G: Nuclear and Particle Physics},
  36(1):015109, 2008.

\bibitem{CANTO200951}
L.F. Canto, P.R.S. Gomes, J.~Lubian, L.C. Chamon, and E.~Crema.
\newblock Dynamic effects of breakup on fusion reactions of weakly bound
  nuclei.
\newblock {\em Nuc. Phys. A}, 821(1):51 -- 71, 2009.

\bibitem{SHORTO200977}
J.M.B. Shorto, P.R.S. Gomes, J.~Lubian, L.F. Canto, S.~Mukherjee, and L.C.
  Chamon.
\newblock Reaction functions for weakly bound systems.
\newblock {\em Phys. Lett. B}, 678(1):77 -- 81, 2009.

\bibitem{redxsec}
P.~R.~S. Gomes, J.~Lubian, I.~Padron, and R.~M. Anjos.
\newblock Uncertainties in the comparison of fusion and reaction cross sections
  of different systems involving weakly bound nuclei.
\newblock {\em Phys. Rev. C}, 71:017601, 2005.

\bibitem{Tho88}
Ian~J. Thompson.
\newblock Coupled reaction channels calculations in nuclear physics.
\newblock {\em Computer Physics Reports}, 7(4):167 -- 212, 1988.

\bibitem{Sat79}
G.R. Satchler and W.G. Love.
\newblock Folding model potentials from realistic interactions for heavy-ion
  scattering.
\newblock {\em Physics Reports}, 55(3):183 -- 254, 1979.

\bibitem{Car80}
L.S. Cardman, J.W. Lightbody, S.~Penner, S.P. Fivozinsky, X.K. Maruyama, W.P.
  Trower, and S.E. Williamson.
\newblock The charge distribution of 12c.
\newblock {\em Physics Letters B}, 91(2):203 -- 206, 1980.

\bibitem{Lah82}
H.~{De Vries}, C.W. {De Jager}, and C.~{De Vries}.
\newblock Nuclear charge-density-distribution parameters from elastic electron
  scattering.
\newblock {\em Atomic Data and Nuclear Data Tables}, 36(3):495 -- 536, 1987.

\bibitem{Hal73}
P.~L. Hallowell, W.~Bertozzi, J.~Heisenberg, S.~Kowalski, X.~Maruyama, C.~P.
  Sargent, W.~Turchinetz, C.~F. Williamson, S.~P. Fivozinsky, J.~W. Lightbody,
  and S.~Penner.
\newblock Electron scattering from $^{19}\mathrm{F}$ and $^{40}\mathrm{Ca}$.
\newblock {\em Phys. Rev. C}, 7:1396--1409, Apr 1973.

\bibitem{RIPL-3}
R.~Capote, M.~Herman, P.~ObloÅ¾inskÃ½, P.G. Young, S.~Goriely, T.~Belgya,
  A.V. Ignatyuk, A.J. Koning, S.~Hilaire, V.A. Plujko, M.~Avrigeanu,
  O.~Bersillon, M.B. Chadwick, T.~Fukahori, Zhigang Ge, Yinlu Han, S.~Kailas,
  J.~Kopecky, V.M. Maslov, G.~Reffo, M.~Sin, E.Sh. Soukhovitskii, and P.~Talou.
\newblock Ripl â€“ reference input parameter library for calculation of
  nuclear reactions and nuclear data evaluations.
\newblock {\em Nuclear Data Sheets}, 110(12):3107 -- 3214, 2009.
\newblock Special Issue on Nuclear Reaction Data.

\bibitem{Coo82}
J.~Cook.
\newblock Dfpot - a program for the calculation of double folded potentials.
\newblock {\em Computer Physics Communications}, 25(2):125 -- 139, 1982.

\bibitem{Wong}
C.~Y. Wong.
\newblock Interaction barrier in charged-particle nuclear reactions.
\newblock {\em Phys. Rev. Lett.}, 31:766--769, 1973.

\bibitem{Zha03}
Zhang Hu-Yong, Shen Wen-Qing, Ren Zhong-Zhou, Ma~Yu-Gang, Chen Jin-Gen, Cai
  Xiang-Zhou, Lu~Zhao-Hui, Zhong Chen, Guo Wei, Wei Yi-Bin, Zhou Xing-Fei,
  Ma~Guo-Liang, and Wang Kun.
\newblock Structures of $^{17}$ {F} and $^{17}$ {O}, $^{17}$ {Ne} and $^{17}$
  {N} in the ground state and the first excited state.
\newblock {\em Chinese Physics Letters}, 20(9):1462--1465, aug 2003.

\bibitem{Ber07}
C.A. Bertulani, G.~Cardella, M.~{De Napoli}, G.~Raciti, and E.~Rapisarda.
\newblock Coulomb excitation of unstable nuclei at intermediate energies.
\newblock {\em Physics Letters B}, 650(4):233 -- 238, 2007.

\bibitem{Ram01}
S.~Raman, C.W. Nestor, and P.~Tikkanen.
\newblock Transition probability from the ground to the first-excited 2+ state
  of even-even nuclides.
\newblock {\em Atomic Data and Nuclear Data Tables}, 78(1):1 -- 128, 2001.

\bibitem{Kee15}
N.~Keeley, K.~W. Kemper, and K.~Rusek.
\newblock Strong multistep interference effects in
  $^{12}\mathrm{C}(d,\phantom{\rule{0.16em}{0ex}}p)$ to the $9/{2}_{1}^{+}$
  state in $^{13}\mathrm{C}$.
\newblock {\em Phys. Rev. C}, 92:054618, Nov 2015.

\bibitem{neon17}
K.~Tanaka, M.~Fukuda, M.~Mihara, M.~Takechi, D.~Nishimura, T.~Chinda,
  T.~Sumikama, S.~Kudo, K.~Matsuta, T.~Minamisono, T.~Suzuki, T.~Ohtsubo,
  T.~Izumikawa, S.~Momota, T.~Yamaguchi, T.~Onishi, A.~Ozawa, I.~Tanihata, and
  T.~Zheng.
\newblock Density distribution of $^{17}${Ne} and possible shell-structure
  change in the proton-rich $\mathit{sd}$-shell nuclei.
\newblock {\em Phys. Rev. C}, 82:044309, 2010.

\bibitem{phosphorus27}
{D.Q. Fang}, {W.Q. Shen}, {J. Feng}, {X.Z. Cai}, {H.Y. Zhang}, {Y.G. Ma}, {C.
  Zhong}, {Z.Y. Zhu}, {W.Z. Jiang}, {W.L. Zhan}, {Z.Y. Guo}, {G.Q. Xiao}, {J.S.
  Wang}, {J.Q. Wang}, {J.X. Li}, {M. Wang}, {J.F. Wang}, {Z.J. Ning}, {Q.J.
  Wang}, and {Z.Q. Chen}.
\newblock Evidence for a proton halo in $^{27}${P} through measurements of
  reaction cross-sections at intermediate energies.
\newblock {\em Eur. Phys. J. A}, 12(3):335--339, 2001.

\end{thebibliography}

\end{document}